\documentclass{pic2012}
\def\runauthor{Andreas B.\,Meyer}
\def\shorttitle{Top Quark Production}
\begin{document}

\title{Top Quark Production}

\author{Andreas B.\,Meyer\\
on behalf of the ATLAS, CDF, CMS and D0 Collaborations}

\address{DESY, Notkestr.85, 22603 Hamburg, Germany\\
E-mail: andreas.meyer@desy.de }

\maketitle

\abstracts{Recent measurements of top quark pair and single top production are presented. The results include inclusive cross sections as well as studies of differential distributions. Evidence for single top quark production in association with a W-boson in the final state is reported for the first time. Calculations in perturbative QCD up to approximate next-to-next-to-leading order show very good agreement with the data.}

\section{Introduction} 

The top quark is by far the heaviest known elementary particle. Due to its large mass the top quark decays within $5\cdot 10^{-25}$s, before hadronisation, and thus gives direct access to its properties such as spin and charge.
With its large mass, the top quark plays a crucial role in electroweak loop corrections, providing indirect constraints on the mass of the Higgs boson. Not least, top quark measurements provide important input to QCD calculations. The measurements help discriminate between different perturbative approaches, and have the potential to constrain QCD parameters. Moreover, various scenarios of physics beyond the Standard Model expect the top quark to couple to new particles. In many super-symmetric models the super-symmetric partner of the top quark is expected to be relatively light, such that it could be produced at the LHC. Experimentally, Standard Model top quark processes are a dominant background to many searches for new physics. 

The top quark was discovered in 1995 at the CDF and D0 experiments at the Tevatron~\cite{cdfobs,d0obs}. Until its shut-down in September 2011, several tens of thousands of top quark events were recorded. The Tevatron collaborations CDF and D0 are now finalizing the measurements with the full statistics. At the LHC at CERN, top quark events are produced at significantly larger rates, and to-date several million top quark events have been recorded by the ATLAS and CMS experiments. The large amount of data gives rise to a wealth of new results and continuous updates. 
In this report an overview of recent measurements of top quark production cross sections is given. Measurements of top quark angular distributions, and properties are reported in~\cite{tony,yvonne}. A recent review article can be found in~\cite{schilling}.

\section{Theoretical Status} 

\begin{figure}[htb]
\centerline{
\epsfxsize=2.2in\epsfbox{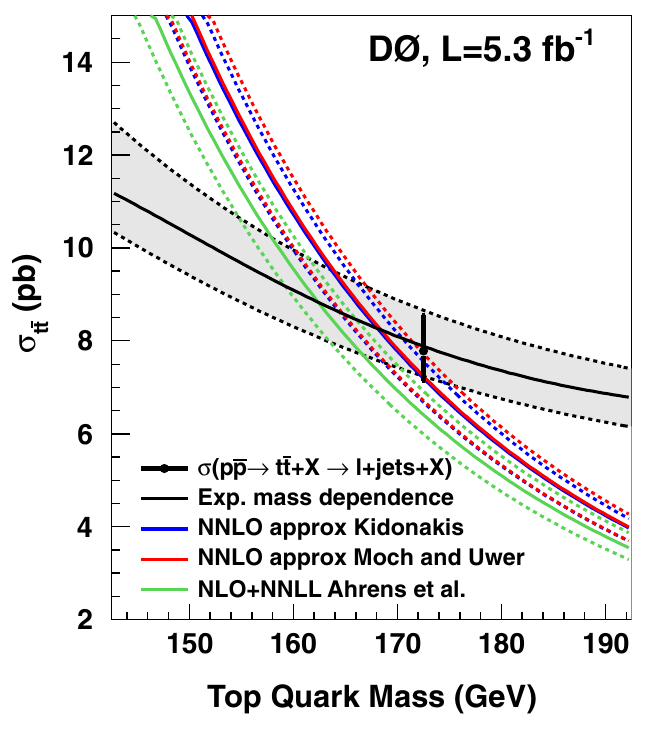} 
\epsfxsize=2.6in\epsfbox{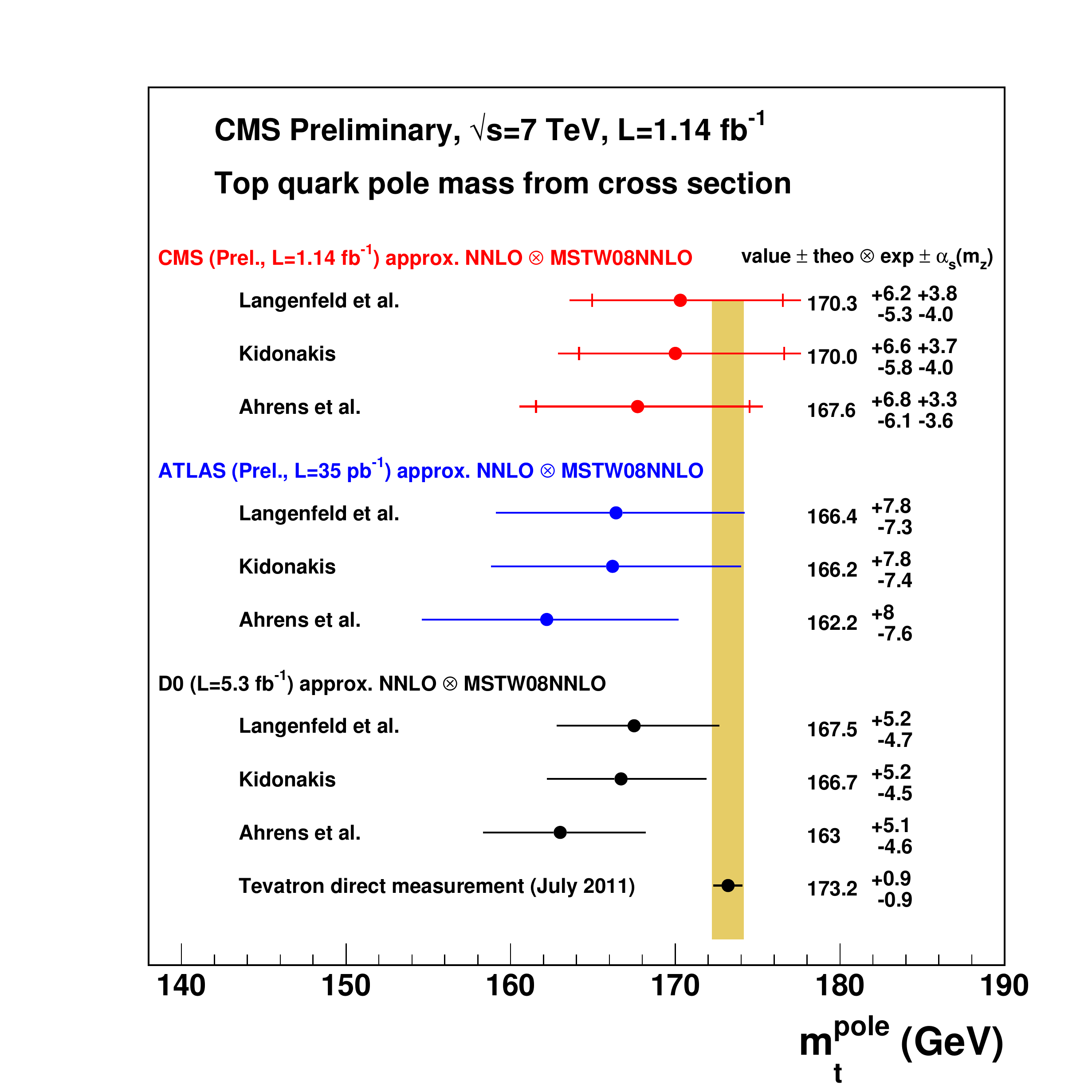}
}
\caption[*]{Left: Dependence of the top quark pair cross section measurement by D0, as well as of various theory calculations, on the mass of the top quark~\cite{d0-prd84-012008}. Right: Summary of indirect measurements of m(pole)~\cite{cms-top-11-008}.}
\label{fig:pole-mass}
\end{figure}

At LHC energies, top quark pair production proceeds dominantly through gluon-gluon fusion. This is in contrast to the Tevatron where the dominant production process is $q\bar{q}$ annihilation at larger values of Bjorken-$x$.
Several groups have performed calculations of the top quark pair production cross section up to approximate next-to-next-to-leading order in perturbative QCD. Most recently, full next-to-next-to-leading order calculations were achieved for the $q\bar{q}$ and $gq$ initial states~\cite{mitov}. The resulting cross section is consistent with next-to-leading order calculations, but the scale uncertainties are significantly reduced, and are now of order 2\%. Calculations for $gg$ are expected very soon. A recent overview of the various available QCD calculations is available here~\cite{beneke}.

Up-to-date Monte Carlo event generators implement the production process at matrix element level up to next-to-leading order~\cite{powheg,mcatnlo}. In tree-level event generators, matrix elements are implemented beyond leading order, containing up to 3 or more hard final state partons~\cite{madgraph,alpgen,sherpa}. In these generators the hadronisation and modeling of the full event final state is generally achieved using parton showers as provided by PYTHIA~\cite{pythia} or HERWIG~\cite{herwig}.

In fixed-order perturbative top quark pair cross section calculations, a theoretically well-defined top quark mass (pole mass or $\overline{MS}$ mass) is used. This is in contrast to calculations as implemented in Monte Carlo generators where the top quark mass does not correspond to a specific renormalization scheme, and soft interactions and parton showers are used to simulate the hadronic final state. The dependence of the cross section prediction on the top quark mass can be exploited to extract the theoretically well-defined pole mass. Such studies have been performed by the D0, CMS and ATLAS experiments, yielding results as depicted in Figure~\ref{fig:pole-mass}~\cite{d0-prd84-012008,d0-polemass,cms-top-11-008,atlas-conf-2011-054}.

\section{Experimental Signatures and Systematics} 

In the Standard Model top quarks are expected to decay to almost 100\% into a $W$ boson and a $b$-quark. Top quark pair events are thus characterized by the presence of two $b$-jets, which can experimentally be tagged, i.e.\,identified, using decay lifetime reconstruction techniques. Top quark pair events are classified experimentally based on the decays of the two $W$ bosons. In the fully hadronic decay channel both $W$ bosons decay into quarks. This final state has the largest branching ratio, but is also most difficult to identify and reconstruct experimentally. The channels where one or both of the $W$ bosons decay into leptons are simpler to measure, and are referred to as $\ell$+jets or dilepton channel, respectively. Triggering and reconstruction details depend on the flavour of the charged leptons (electron, muon or $\tau$). Analyses with $\tau$-leptons use the decays of the $\tau$ into low multiplicity jets. In leptonic $W$ boson decays large momentum neutrinos are produced which escape detection, thus giving rise to large missing transverse momentum. 

\section{Inclusive Cross Section Measurements}

\begin{figure}[htb]
\centerline{
\epsfxsize=2.7in\epsfysize=2.5in\epsfbox{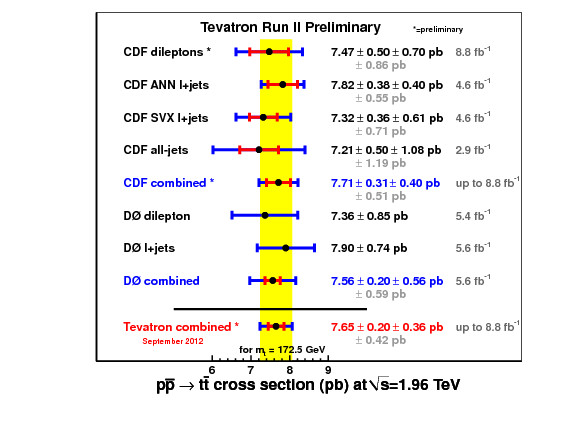}
\epsfxsize=2.5in\epsfysize=2.5in\epsfbox{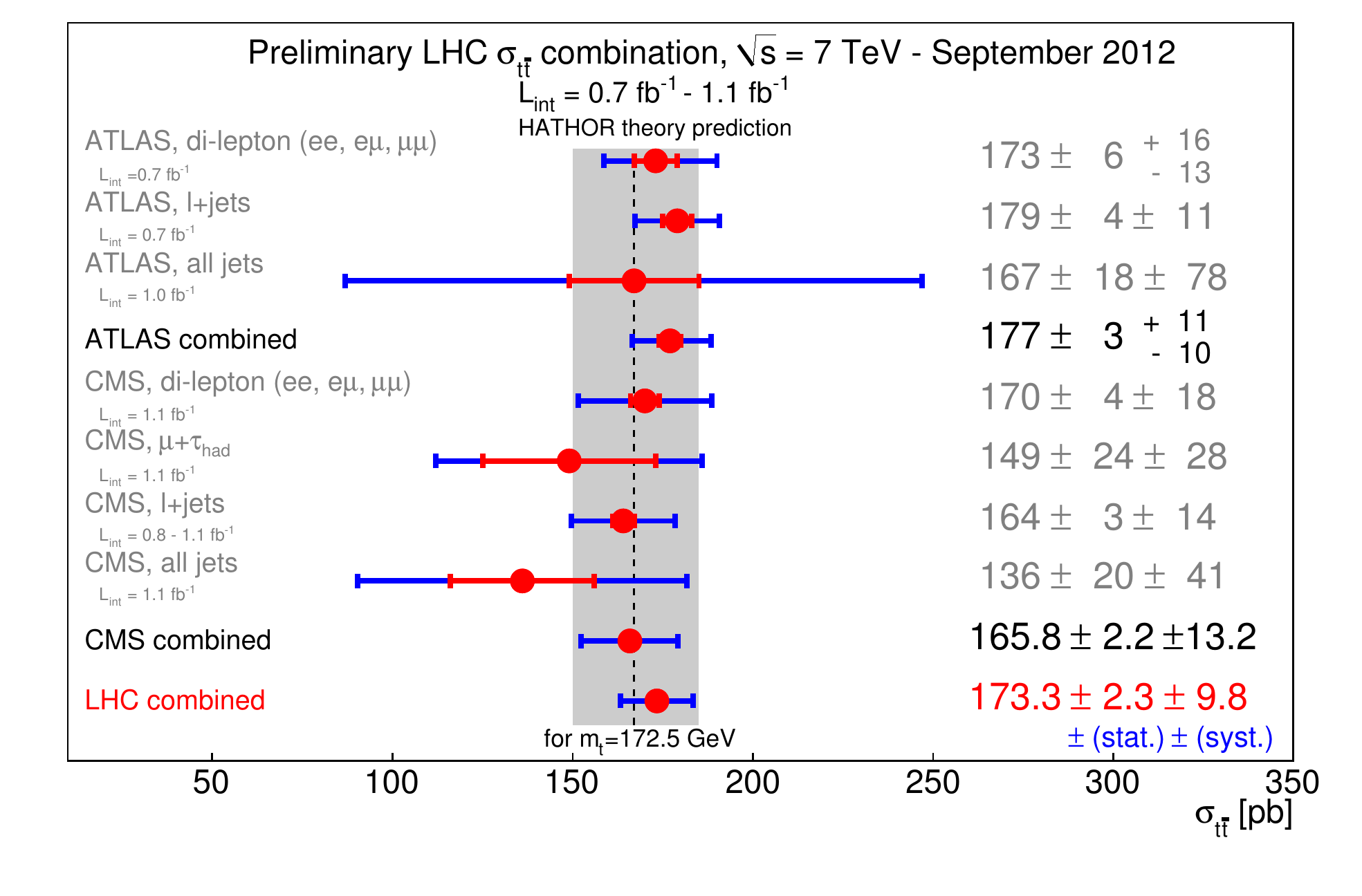}
}
\caption[*]{Overview and Combination of top quark pair cross section measurements at the Tevatron (left)~\cite{cdfweb} and at the LHC (right)~\cite{atlas-conf-2012-134}.}
\label{fig:comb}
\end{figure}

\begin{figure}[htb]
\centerline{
\epsfxsize=2.3in\epsfysize=1.8in\epsfbox{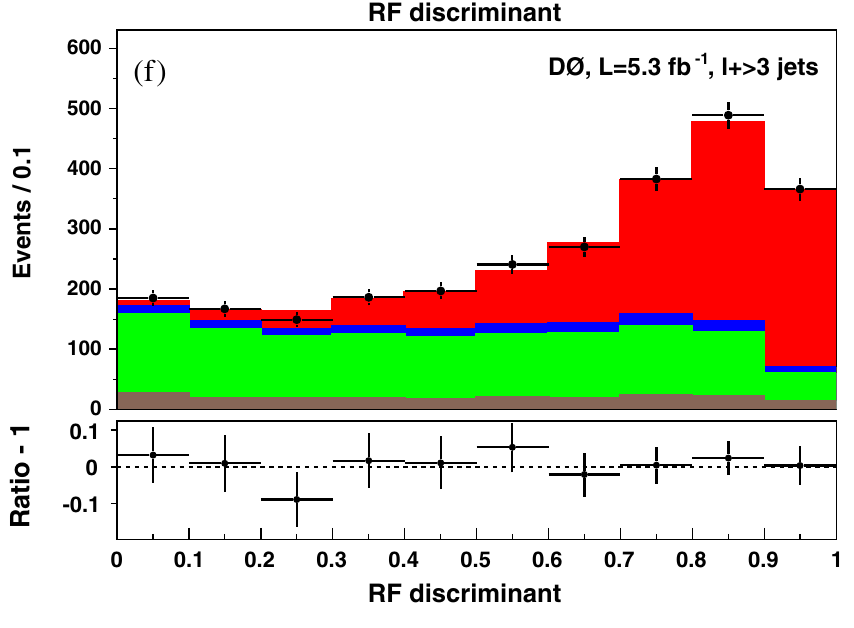} ~
\epsfxsize=2.5in\epsfysize=1.8in\epsfbox{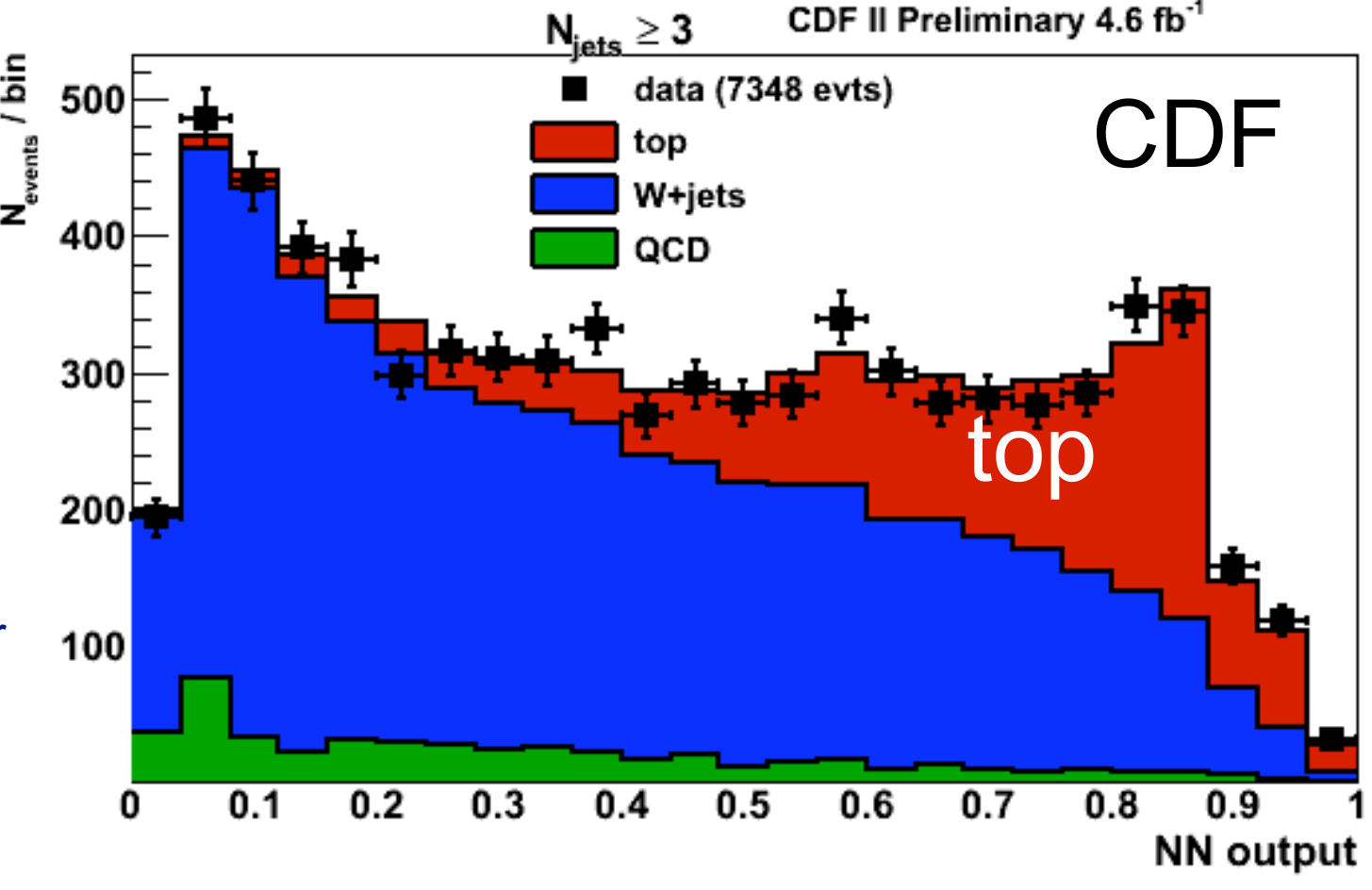}
}
\caption[*]{Output distributions from multivariate analyses from D0 (left)~\cite{d0-prd84-012008} and CDF (right)~\cite{cdf-prl105-012001}.}
\label{fig:tev-incl}
\end{figure}

Measurements of the inclusive top quark pair cross section have been performed using all decay channels (except the one with two $\tau$-leptons in the final state). The combination of the most recent measurements at the Tevatron, using data of an integrated luminosity of up to 8.8 fb$^{-1}$, yields a cross section of $7.65 \pm 0.20 \pm 0.36$ pb, with a relative total uncertainty of 5.5\%. An overview of the various analyses is given in Figure~\ref{fig:comb}(left). The two single most precise results from D0 and CDF are obtained in the $\ell$+jets channel~\cite{d0-prd84-012008,cdf-prl105-012001}. Both results achieve optimal separation of the top quark signal from the background by combining results from analyses based on $b$-tagging with multivariate techniques. In Figure~\ref{fig:tev-incl} the output distributions from the multivariate algorithms are shown. The D0 measurement, $\sigma_{t\bar{t}}=7.78^{+0.77}_{-0.64}$ pb, is obtained from a simultaneous fit of signal and background distributions. In the fit, a scale factor for the contribution from events with W-bosons and heavy quark jets is also determined. The dominant systematics originate from uncertainties on the integrated luminosity as well as jet-identification. The result from CDF is $\sigma_{t\bar{t}}=7.70\pm 0.54$ pb. Here, the luminosity uncertainty is minimized by normalizing the measurement to the measured number of Drell-Yan events where $Z^0$ decays into two leptons. For the latter measurement, the same triggers and lepton identification cuts are applied, such that the uncertainties are further reduced. The results are in good agreement with each other and with theory calculations at NNLO+NNLL which yield $\sigma_{t\bar{t}}=7.24^{+0.15}_{-0.24}(scale)^{+0.18}_{-0.12}(PDF)$ pb~\cite{mitov}. In this calculation, performed for a top quark mass of 172.5 GeV, the contribution $q\bar{q}\rightarrow t\bar{t}$ is calculated to full NNLO~\cite{mitov}. Both collaborations D0 and CDF are now in the process of finalizing the top quark analyses using their full datasets, corresponding to about 10 fb$^{-1}$. 

\begin{figure}[htb]
\centerline{
\epsfxsize=2.5in\epsfysize=2.3in\epsfbox{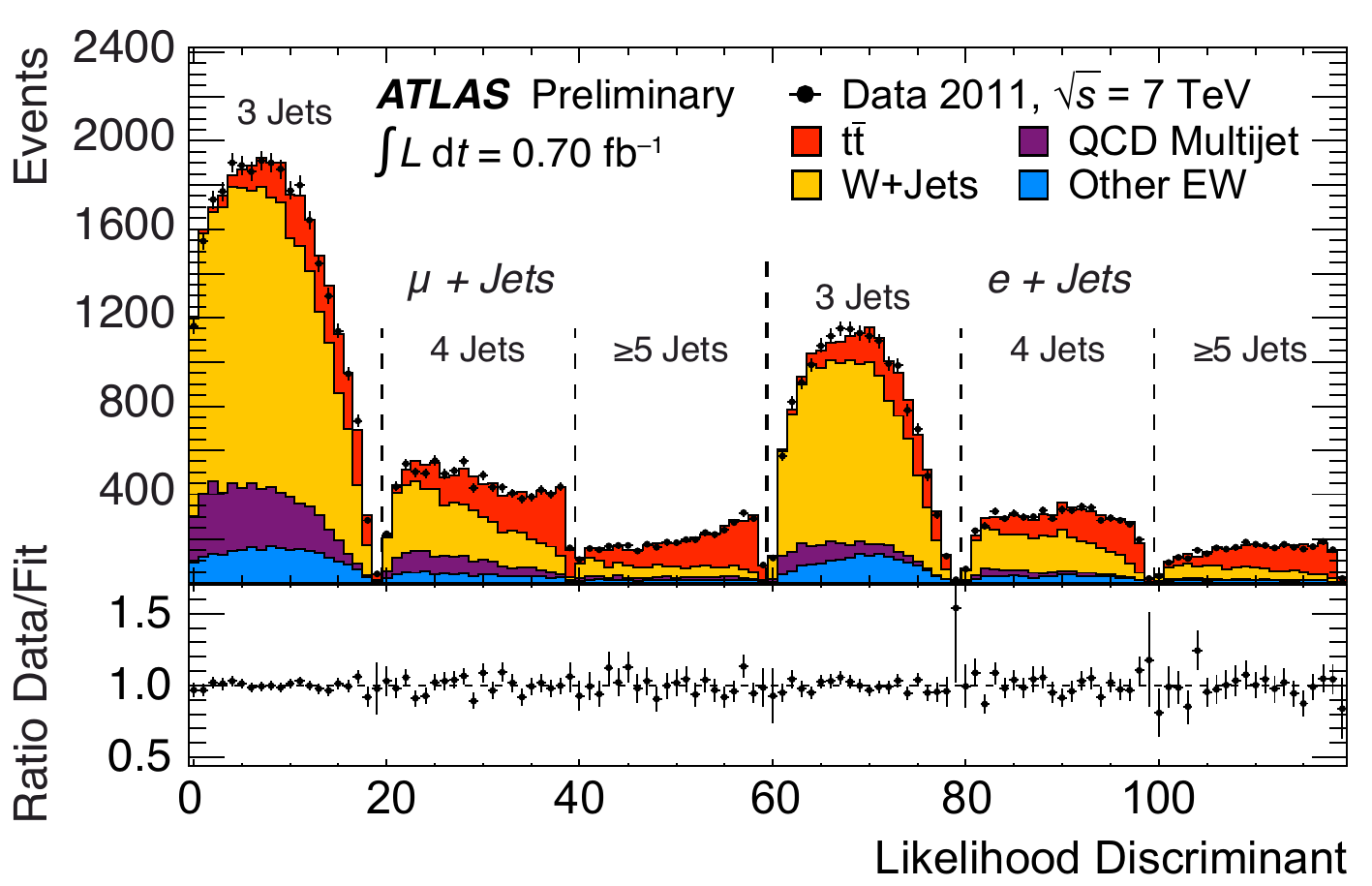} 
\epsfxsize=2.5in\epsfysize=2.3in\epsfbox{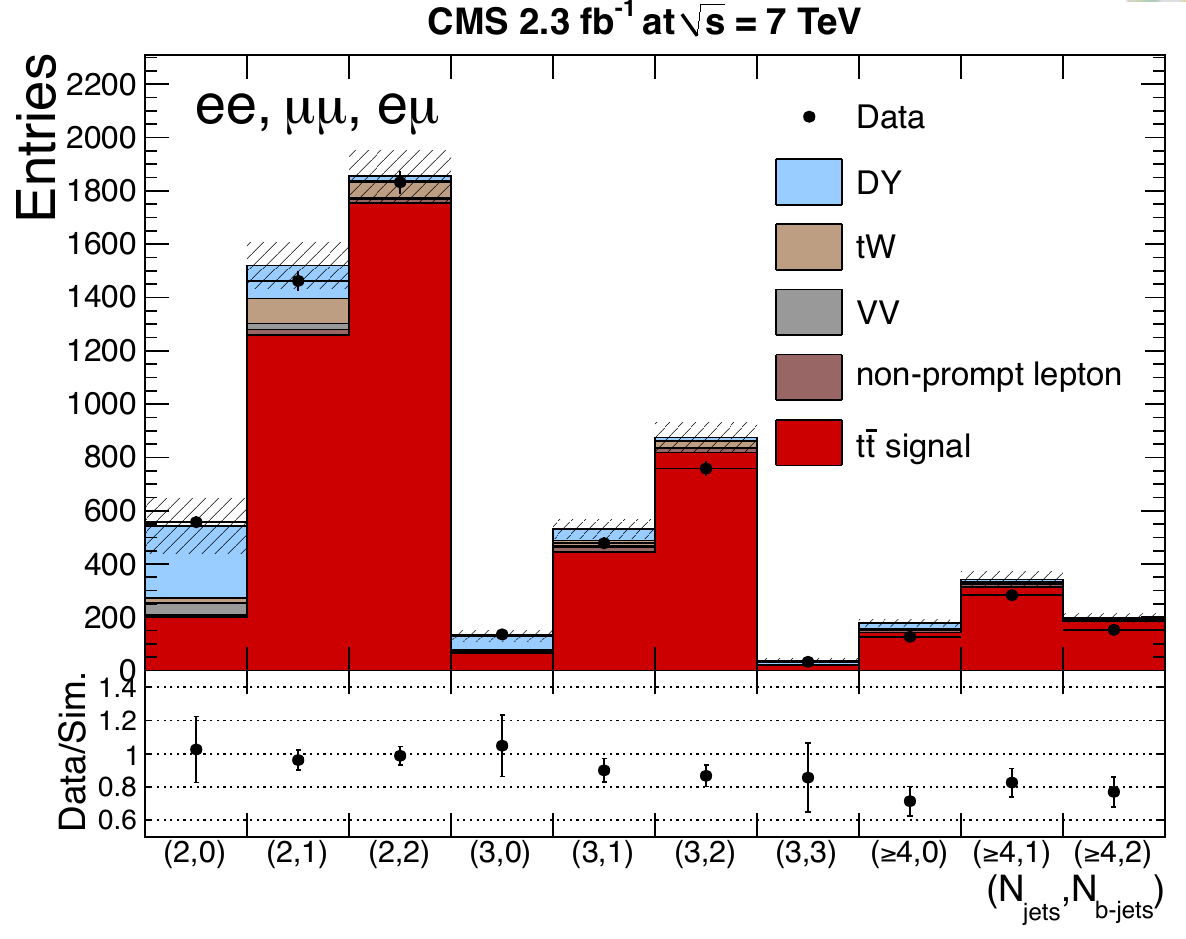}
}
\caption[*]{Distributions of top quark pair cross section measurements from ATLAS(left)~\cite{atlas-conf-2011-121} and CMS(right)~\cite{cms-top-11-005}.}
\label{fig:lhc-incl}
\end{figure}

A compilation of measurements at the LHC using the dataset recorded in the year 2011 at a $pp$ centre-of-mass energy of 7 TeV are presented in Figure~\ref{fig:comb}(right)~\cite{atlas-conf-2012-134}.
The most precise results from the LHC were achieved in the $\ell$+jets channel (ATLAS)~\cite{atlas-conf-2011-121} and the dilepton channel (CMS)~\cite{cms-top-11-005}. The latter is not yet included in Figure~\ref{fig:comb}.
The ATLAS analysis is based on a template fit to a likelihood discriminant using event kinematic information as input (Figure~\ref{fig:lhc-incl}(left)). The measured cross section is $\sigma_{t\bar{t}}=179.0\pm3.9(stat.)\pm9.0(syst.)\pm6.6(lumi.)$ pb corresponding to a total relative uncertainty of 6.5\%. Dominant systematic uncertainties arise from the modeling of the signal, the jet energy scale and the lepton identification uncertainties. 
The CMS analysis uses a profile likelihood fit to the 2-dimensional jet and $b$-tag multiplicity distribution (Figure~\ref{fig:lhc-incl}(right)). The measured cross section is $\sigma_{t\bar{t}}=161.9\pm2.5(stat.)\pm5.1(syst.)\pm3.6(lumi.)$ pb corresponding to a total relative uncertainty of 4.2\%. This is the single most precise top quark cross section measurement so far. Dominant systematic uncertainties are related to the jet energy scale and the lepton identification uncertainty. 

\begin{figure}[htb]
\centerline{
\epsfxsize=3.10in\epsfbox{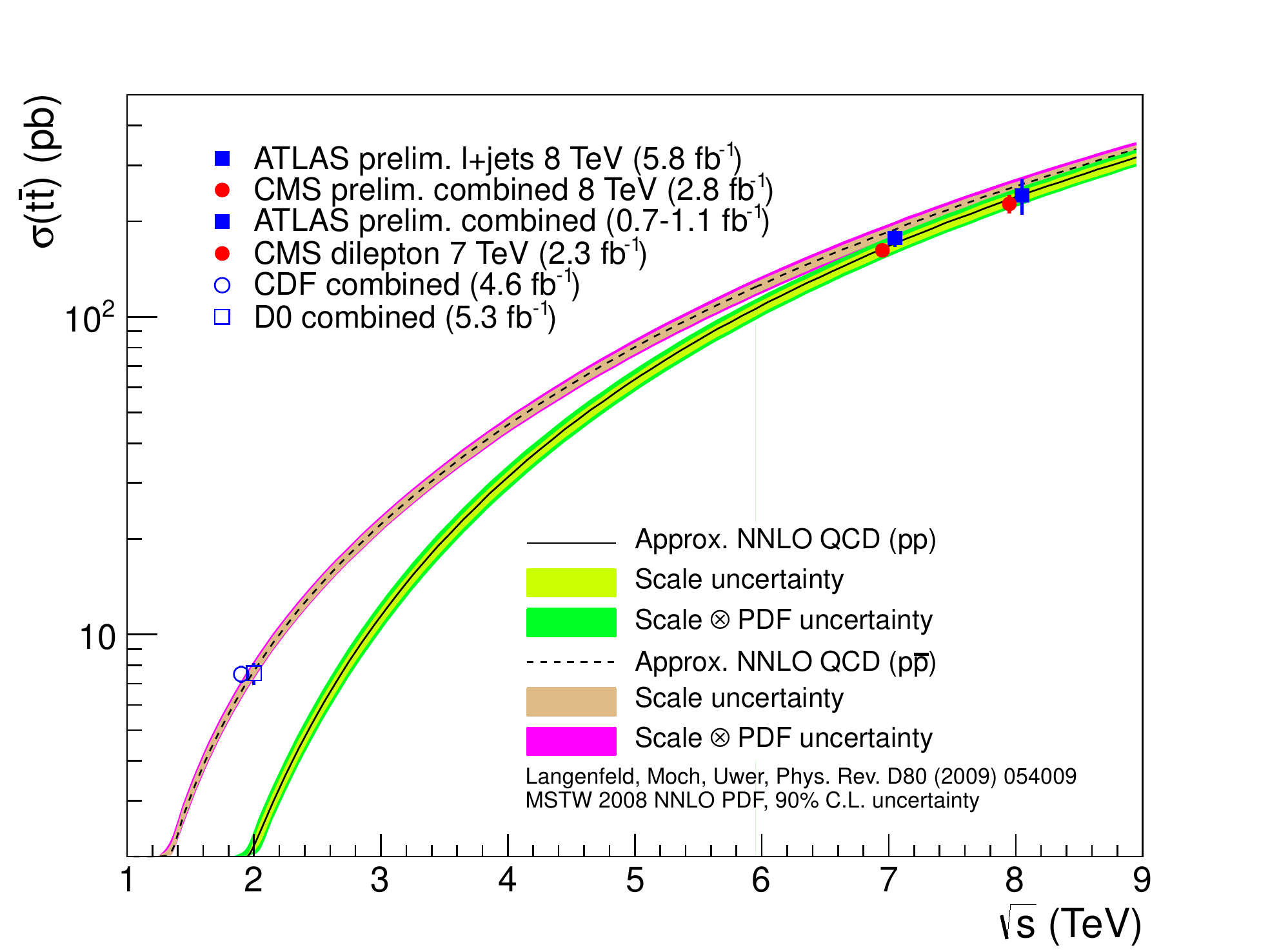}
}
\caption[*]{Top quark pair cross section measurements as a function of the $pp$ and $p\bar{p}$ centre-of-mass energy. The bands represent result from calculations in perturbative QCD up to approximate NNLO}
\label{fig:sqrts}
\end{figure}

Both CMS and ATLAS have presented first results at a $pp$ centre-of-mass energy $\sqrt{s}=8$ TeV~\cite{cms-top-12-006,cms-top-12-007,atlas-conf-2012-149}.
A summary of the inclusive top quark pair cross section measurements at the Tevatron and the LHC is shown in Figure~\ref{fig:sqrts}. Good agreement with QCD calculations up to approximate NNLO is observed.

\section{Differential Distributions}

Additional information about the physics of top quark production can be gained from measurements of differential distributions. The measurement of kinematic top quark distributions does not only probe QCD predictions and provide input to an improved choice of QCD model and scale parameters. Differential distributions also have the potential to constrain the parton distribution functions of gluons at large x. Moreover, the distributions are sensitive to possible new physics which are especially expected to occur at high $t\bar{t}$ invariant masses, for example decays of massive Z-like bosons into top quark pairs.

The kinematic properties of the top quark pair are determined from the four-momenta of all final-state objects by means of kinematic reconstruction algorithms. In the $\ell$+jets channels constrained kinematic fitting algorithms are applied to obtain the kinematics of both top quarks. In the dilepton channels, due to the presence of two neutrinos, the kinematic reconstruction is underconstrained, even after imposing the full set of possible kinematic constraints such as that of the W-boson invariant mass of 80.4 GeV, the equality of the top quark and antiquark masses and assuming that the missing energy originates solely from the neutrinos in the event. Ambiguities between several solutions are resolved by prioritization e.g.\,by use of the expected neutrino energy distribution.

\begin{figure}[htb]
\centerline{ 
\epsfxsize=2.5in\epsfysize=2.5in\epsfbox{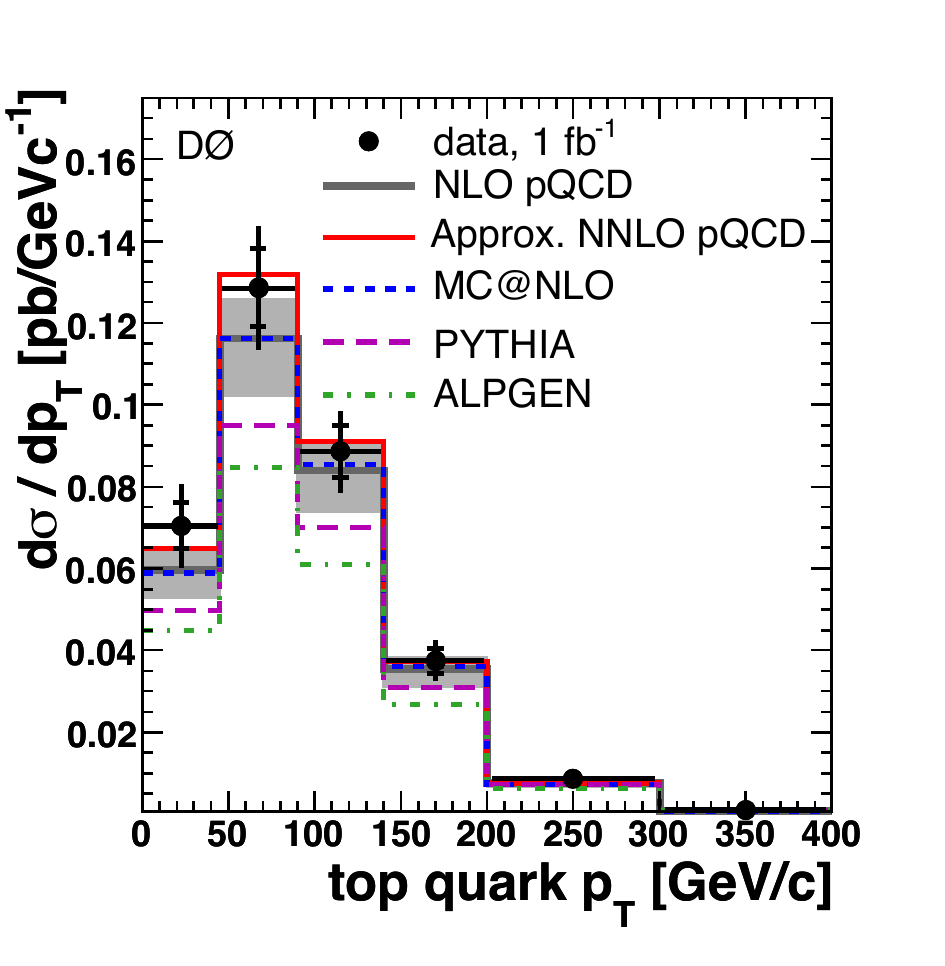} ~
\epsfxsize=2.5in\epsfysize=2.5in\epsfbox{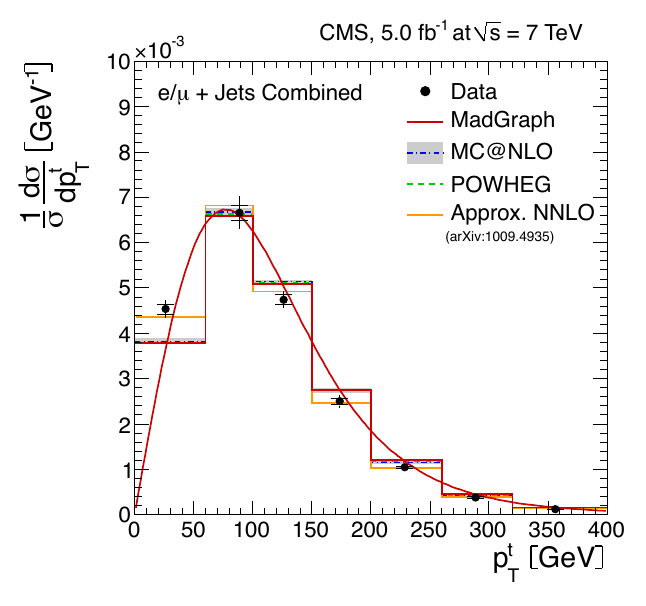}}
\caption[*]{Distributions of the top quark transverse momentum in top quark pair events at the Tevatron (D0) (left)~\cite{d0-plb-693-515} and the LHC (right)~\cite{cms-top-11-013}.}
\label{fig:diff-pttop}
\end{figure}

First differential measurements of $t\bar{t}$ cross sections were performed by CDF and D0~\cite{cdf-PRL-102-222003,Ahrens,d0-plb-693-515}.
In general, good agreement of theoretical calculations with the data was observed. In Figure~\ref{fig:diff-pttop} (left) the differential cross section as a function of transverse momentum of the top quark is shown as measured by D0~\cite{d0-plb-693-515}. Calculations including higher order corrections (up to approximate NNLO) are found to give an improved description of the data, especially w.r.t.\,their normalization.

At the LHC, the large $t\bar{t}$ production rate leads to a substantial reduction of the statistical uncertainties in each bin, and in turn helps reduce systematics. In Figure~\ref{fig:diff-pttop}(right) the measurement of the transverse momentum distribution of top quarks by CMS is presented~\cite{cms-top-11-013}. The measurement makes use of the full statistics accumulated in the year 2011 at 7 TeV and is limited by systematic uncertainties. Several models are confronted with the data. They all give a good description of the data. A yet improved description is achieved by the prediction to approximate NNLO.

\begin{figure}[htb]
\centerline{
\epsfxsize=2.5in\epsfysize=2.3in\epsfbox{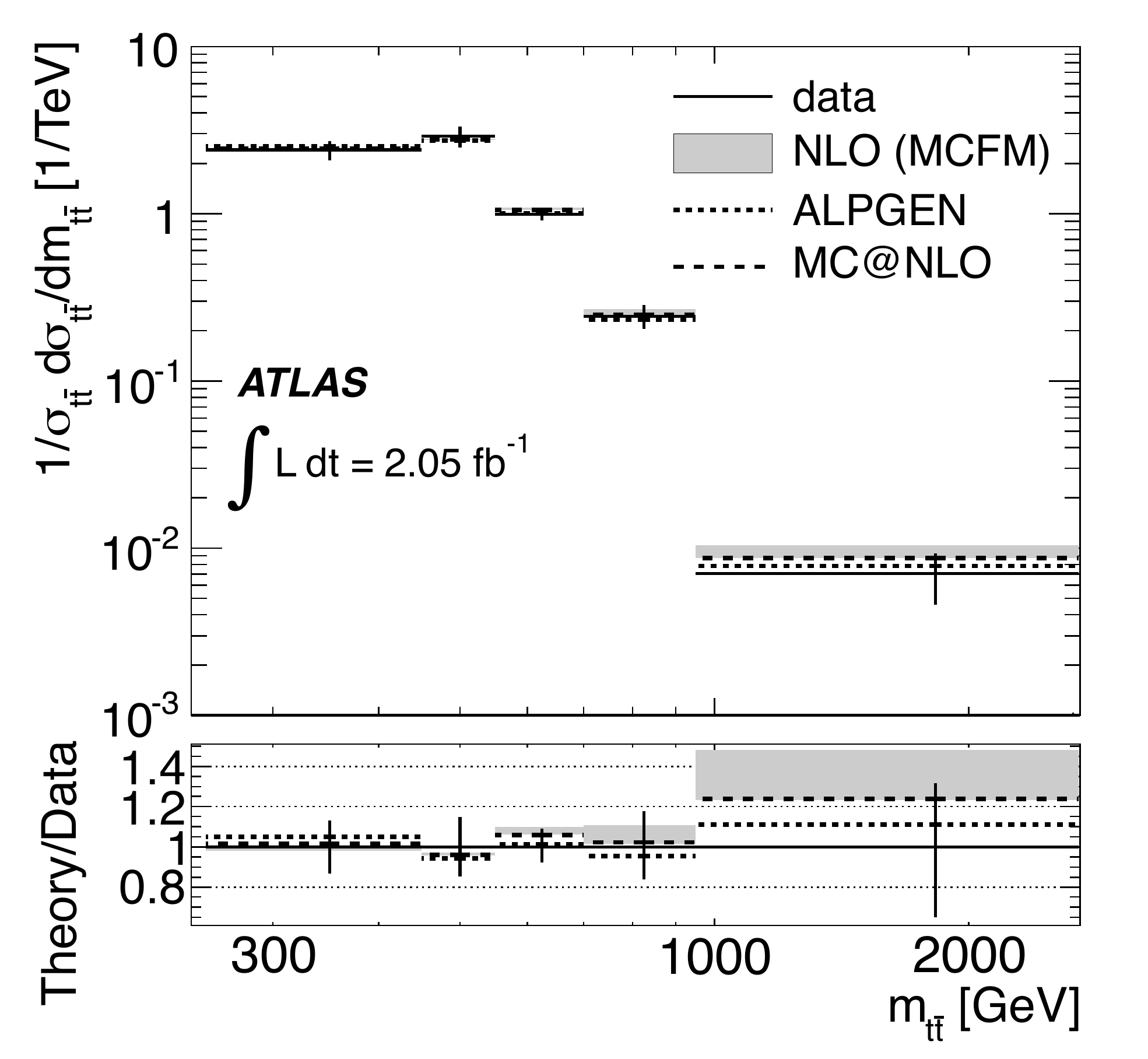} ~
\epsfxsize=2.5in\epsfysize=2.45in\epsfbox{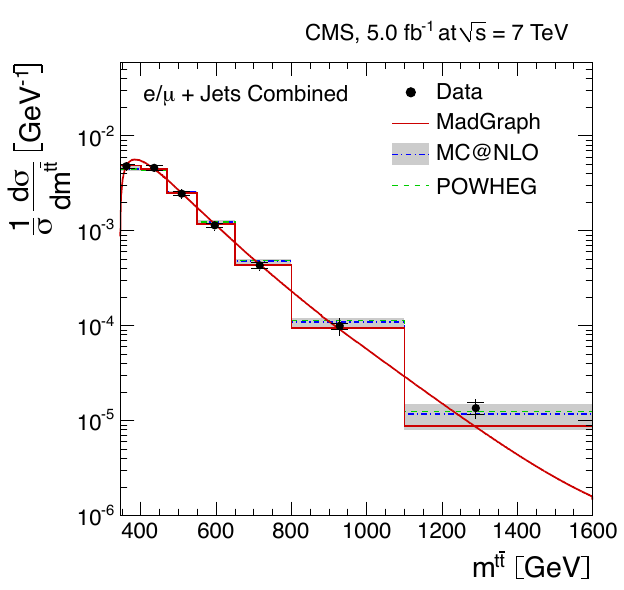}
}
\caption[*]{Distributions of the invariant mass of the $t\bar{t}$ system from ATLAS(left)~\cite{atlas-diff} and from CMS(right)~\cite{cms-top-11-013}.}
\label{fig:diff-mtt}
\end{figure}

A large number of distributions of the top quark and the $t\bar{t}$ system, as well as their decay products, has been measured at the LHC~\cite{cms-top-11-013,atlas-diff}. ATLAS and CMS report normalized differential cross sections, i.e.\,shape measurements, in which normalization uncertainties are removed.
In Figure~\ref{fig:diff-mtt} the distributions of the invariant mass of the $t\bar{t}$ system are displayed.
Both the ATLAS and the CMS data are very well described by the various calculations up to an energy scale of more than 1~TeV.

\section{Jet Multiplicity Distributions}

\begin{figure}[htb]
\centerline{
\epsfxsize=2.5in\epsfysize=2.5in\epsfbox{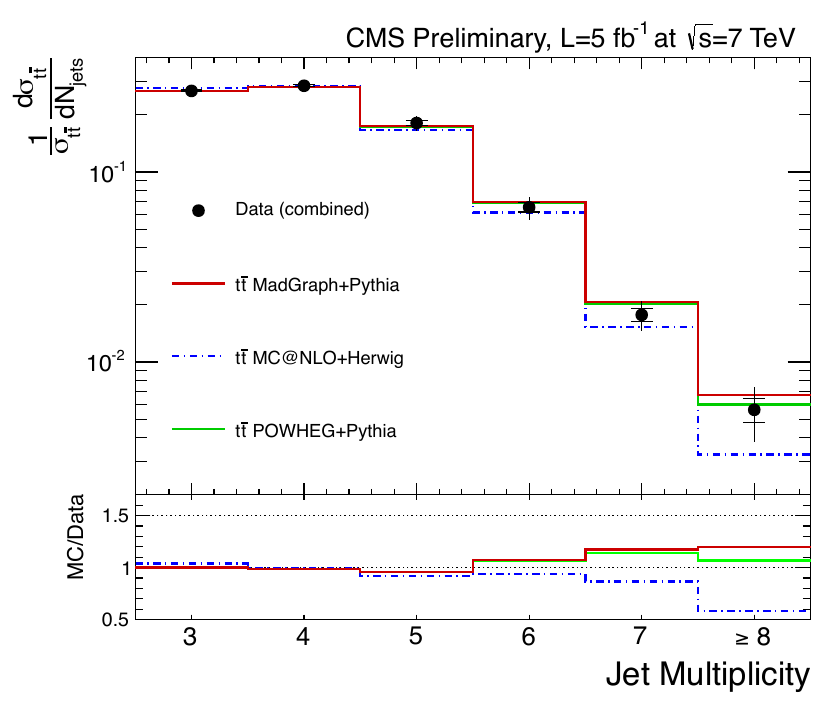}
\epsfxsize=2.5in\epsfysize=2.4in\epsfbox{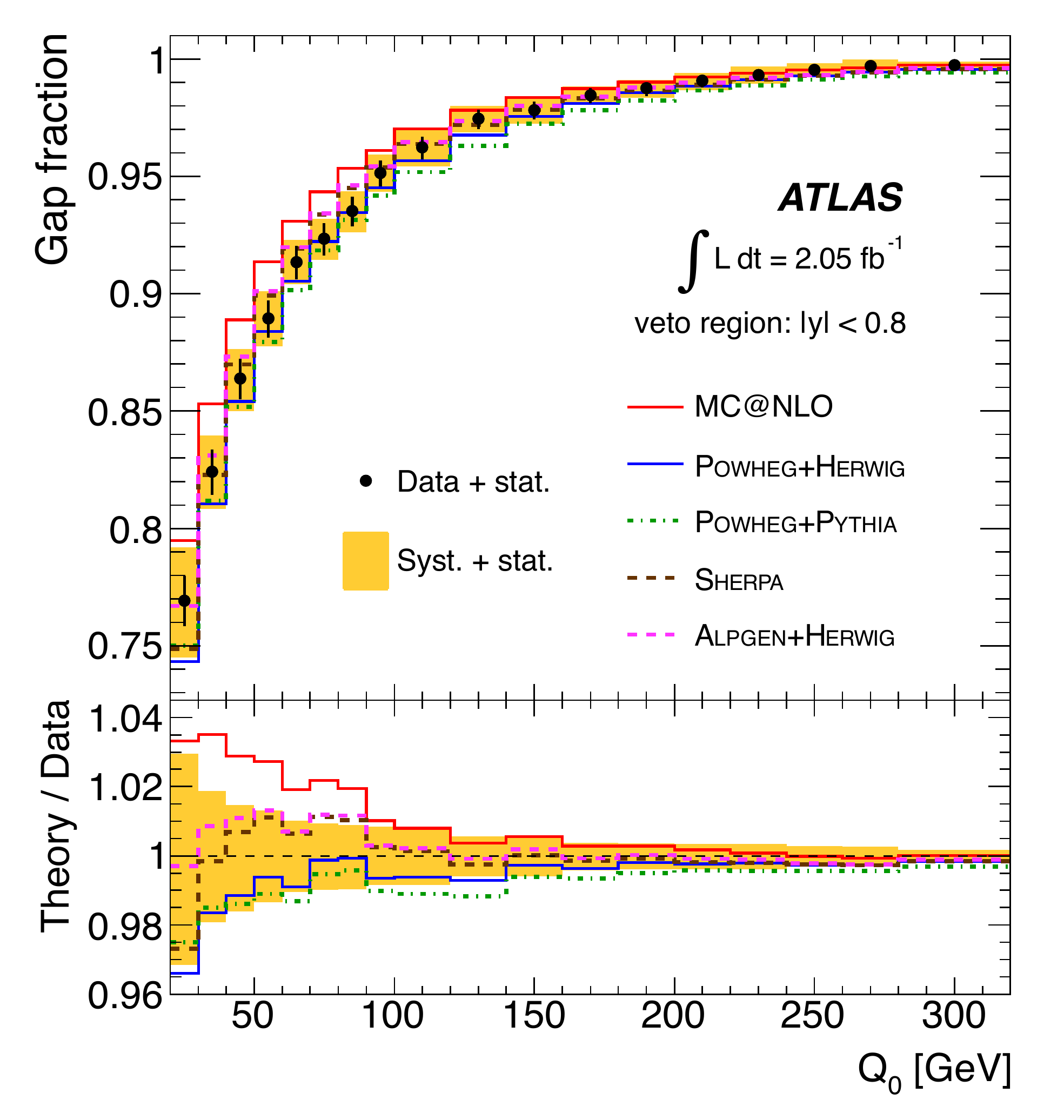}
}
\caption[*]{Normalized differential cross section as a function of jet multiplicity (left)~\cite{cms-top-12-018} and the gap fraction, i.e.\,the fraction of events surviving the jet veto as a function of the transverse momentum of the jet (right)~\cite{atlas-gap}.}
\label{fig:ttjets}
\end{figure}

At LHC energies, the fraction of top quark pair events with additional hard jets in the final state is large, about half of the total number of events~\cite{ATLAS-CONF-2012-083}. For the correct description of events with additional jets, contributions from higher order QCD processes are required which take into account additional radiation in the initial or final state. The understanding of these processes is important not least because multi-jet processes constitute important backgrounds for many new physics searches. In recent measurements from ATLAS and CMS the distributions of jet multiplicities and additional jets due to QCD radiation are studied in detail~\cite{atlas-conf-2011-142,atlas-conf-2012-155,atlas-gap,cms-top-12-018,cms-top-12-023}.
In Figure~\ref{fig:ttjets}(left) the multiplicity distribution of jets for top quark pair events in the $\ell$+jets channel is shown. The data are generally well described by the Monte Carlo generators. Towards large multiplicity the MC@NLO generator interfaced with parton shower from HERWIG is seen to predict significantly less events than MADGRAPH or POWHEG which use PYTHIA to generate the parton showers.
An alternative way of investigating additional activity in the event is to study the gap fraction distribution. Events are vetoed if they contain an additional jet with transverse momentum above threshold in a central rapidity interval. The fraction of events surviving the jet veto, the gap fraction, is presented as a function of this threshold. The gap fraction distribution for jets in the central rapidity range is displayed in  Figure~\ref{fig:ttjets}(right). A qualitatively similar trend is observed as in the multiplicity distribution in that the MC@NLO generator predicts a larger fraction of events that have no jet activity beyond the jets originating directly from the top quark decays. As the data are able to discriminate between the predictions from different models, these results can be used to optimize the choice of models.

\section{Single Top}

Single top quark production, in the Standard Model, is expected to proceed through charged-current electro-weak interactions. Depending on whether the W-boson is time-like, space-like or real, one distinguishes between the $s$-channel, the $t$-channel and the $tW$-channel. In the latter channel, single top quarks are produced in association with a $W$ boson in the final state.

Measurements of single top production constitute a unique test of the electro-weak interactions and quark-flavour dynamics as described by the CKM matrix. The measured single top production cross sections can be used for direct constraints of the CKM matrix element $|V_{tb}|$ and of possible contributions from new physics, e.g.\,arising from feed-down from potential fourth-generation quarks. The ratio of top and anti-top production is sensitive to $b$-quark parton density distributions as well as the ratio of $u$ and $d$ valence quarks. 

Single top production was first discovered at the Tevatron in 2009, much later than top quark pair production due to the significantly larger backgrounds as well as the smaller cross sections of electro-weak single top production in comparison to the strongly produced top quark pairs. The Tevatron analyses are performed for a combination of $t$-channel and $s$-channel, using multi-variate analysis techniques to separate the signals from the background~\cite{d0-prd-84-112001,d0-plb-705-313,cdf-conf-10793}.

Depending on the center-of-mass energy, and the corresponding initial state parton distributions and phase space, the expected cross sections for the three production channels are very different between the Tevatron and the LHC. At the Tevatron, only the $s$ and $t$-channel are expected to have measurable rates of similar order of magnitude, while at the LHC, the $t$ and $tW$ channel have large cross sections, while the $s$-channel is significantly smaller in rate. Experimentally, for trigger and background reasons, single top measurements are performed using the leptonic decays of the W-boson from the top quark.

\begin{figure}[htb]
	
\centerline{
\epsfxsize=2.5in\epsfysize=2.5in\epsfbox{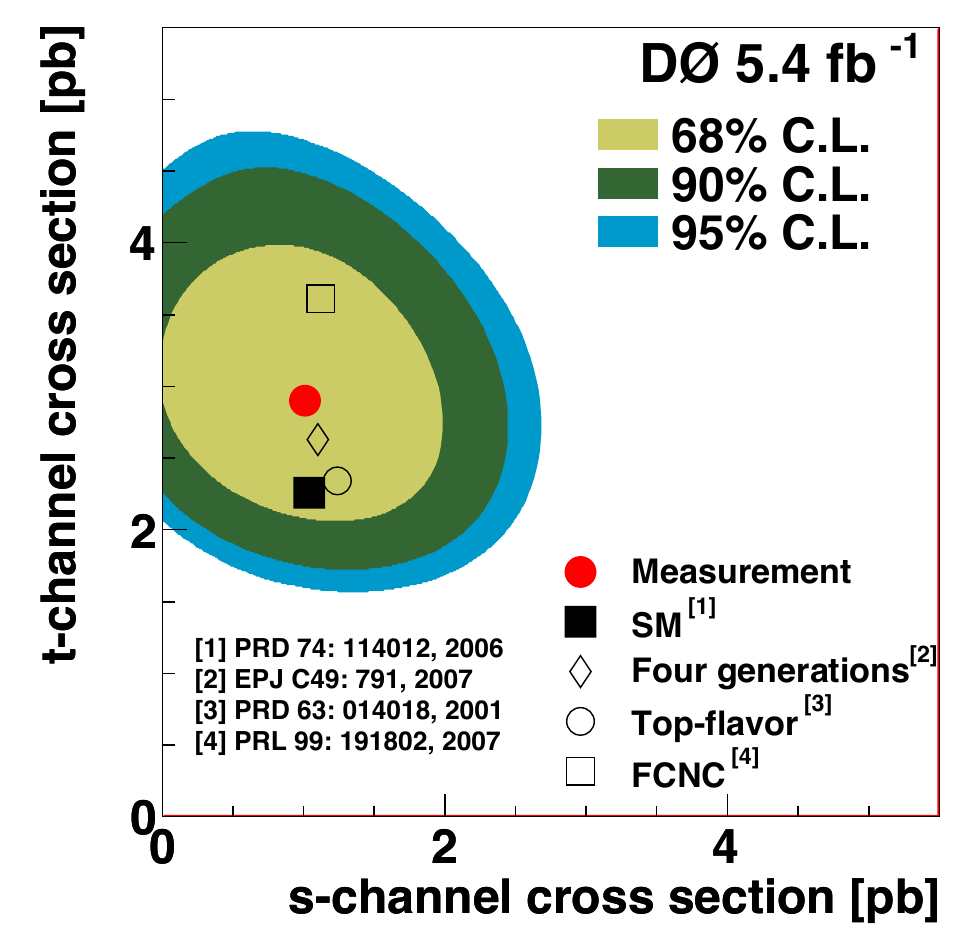} 
\epsfxsize=2.5in\epsfysize=2.5in\epsfbox{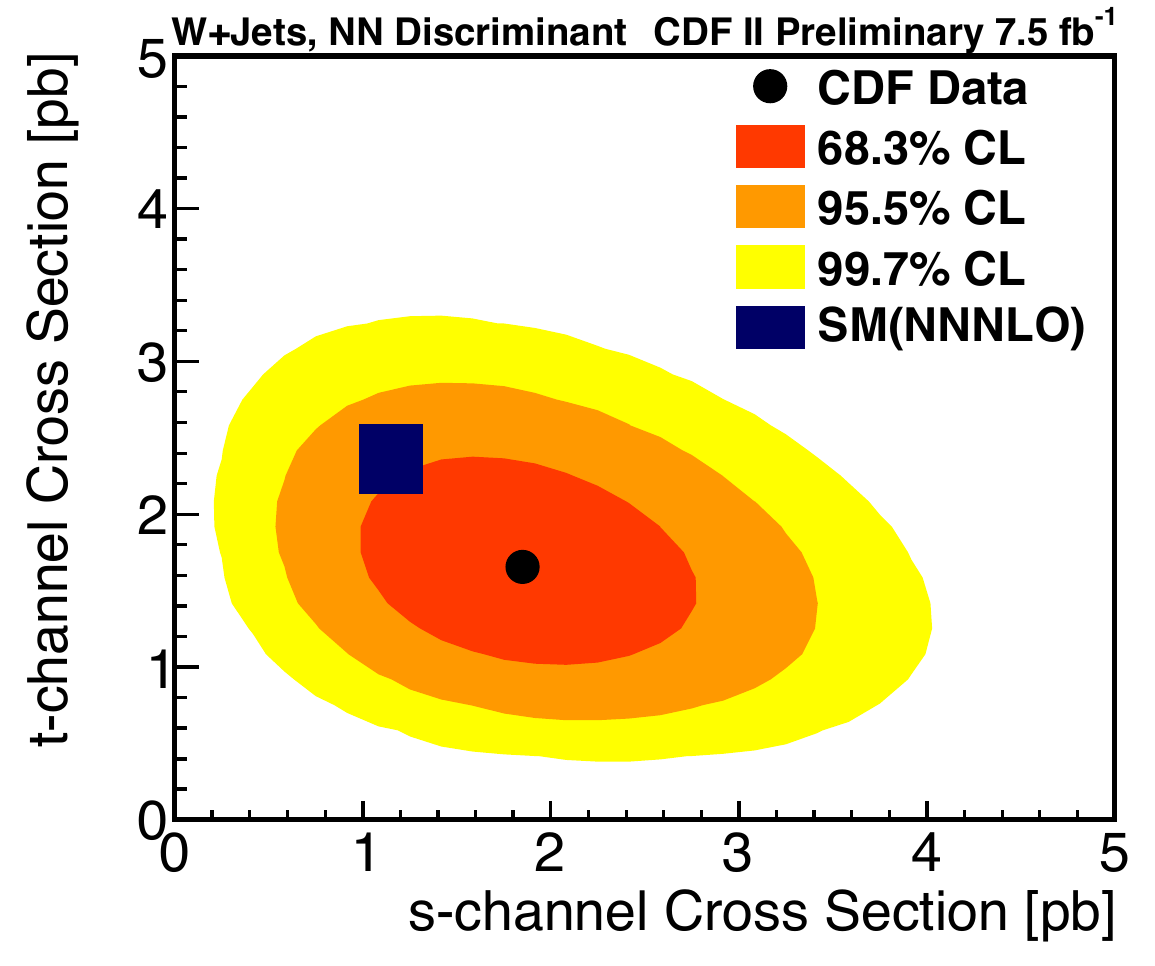}}
\caption[*]{Results of the single top production cross section measurements in the $t$-channel and in the $s$-channel from the D0 Experiment (left)~\cite{d0-plb-705-313} and the CDF Experiment (right)~\cite{cdf-conf-10793}.}
\label{fig:tev-single}
\end{figure}

In the case of D0, a combination of three multi-variate analyses is used. Each MVA method is trained separately for the two single top quark production channels: for the $t$-channel discriminants, with $t$-channel considered signal and $s$-channel treated as a part of the background, and for the $s$-channel discriminants, with $s$-channel considered signal and $t$-channel contributions treated as a part of the background.
For the combined measurement of the sum of $s$- and $t$-channel, the Standard Model prediction for the ratio between $s$- and $t$-channel is used as input.
The results from the three different analyses are combined, yielding a sum of cross sections of $\sigma_{s+t}=3.43^{+0.73}_{-0.74}$ pb. The main systematic uncertainties arise from uncertainties related to the jet energy scale, the $b$-jet identification and the integrated luminosity.
Figure~\ref{fig:tev-single}(left) shows the result of the analysis in which the $t$-channel is treated as signal.

The CDF Experiment performs a multi-variate analysis based on a neural network. For the sum of $s$ and $t$-channel, the Standard Model prediction for the ratio between $s$- and $t$-channel is used as constraint, and the cross section is measured to be $\sigma_{s+t}=3.04^{+0.57}_{-0.53}$ pb.
The result for the case in which the $s$-channel is treated as signal is shown in Figure~\ref{fig:tev-single}(right). Both CDF and D0 establish a clear observation of single top quark production, in good quantitative agreement with the Standard Model prediction. 

At the LHC, the cross section for single top production is significantly larger than at the Tevatron.
At a centre-of-mass energy of 7 TeV, the cross section for $t$-channel production alone is predicted to be $\sigma_t=64^{+3.3}_{-2.6}$ pb~\cite{kidonakist}, more than one third of the top quark pair cross section. Due to the large LHC luminosity and excellent background suppression capabilities, CMS and ATLAS have been able to perform measurements of the $t$-channel and $tW$-channel cross sections, with already very good precision. 

\begin{figure}[htb]
\centerline{
\epsfysize=2.5in\epsfbox{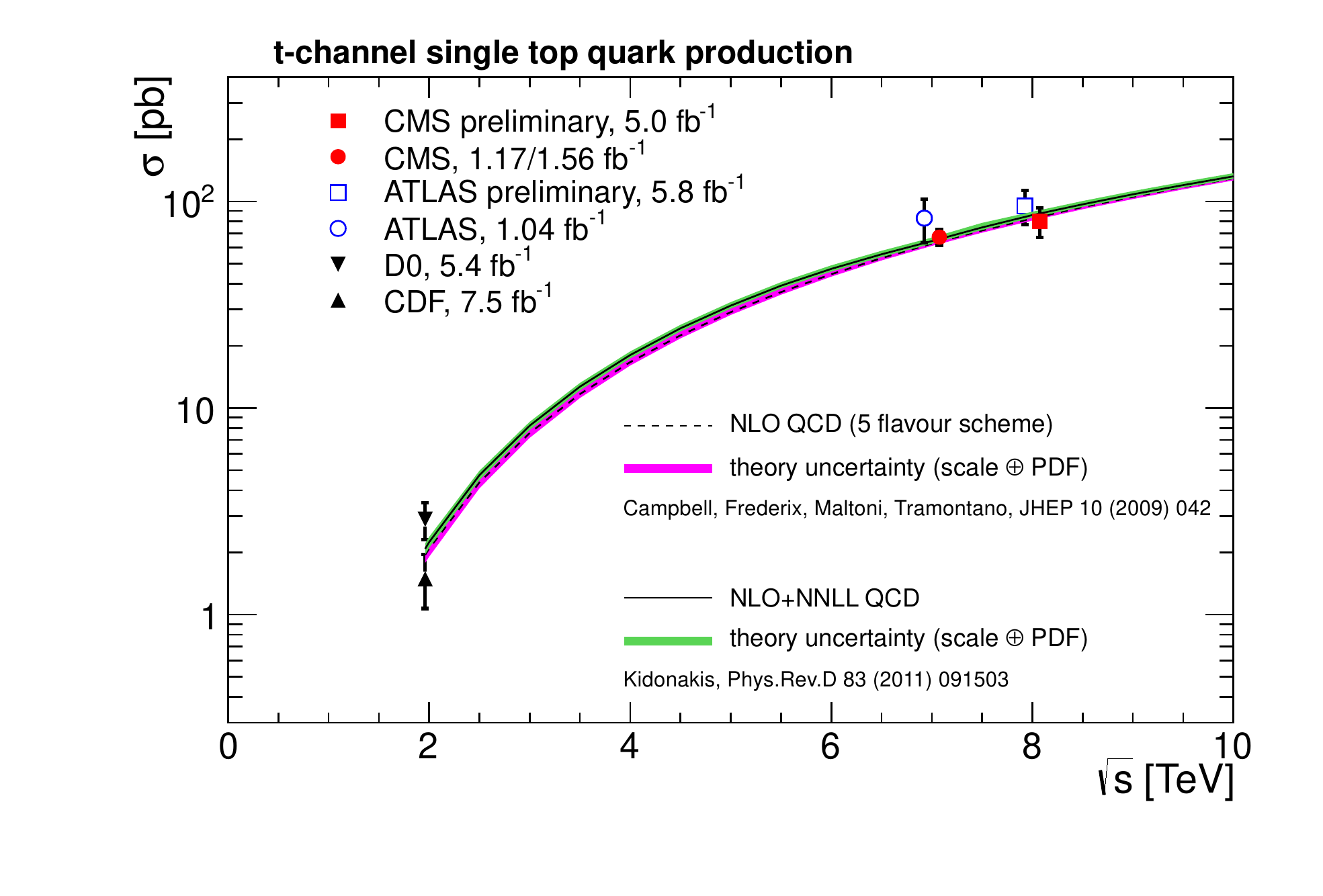} 
}
\caption[*]{Overview of $t$-channel measurements at the Tevatron and the LHC. The measurements are displayed as a function of the center of mass energy.}
\label{fig:singlet-sqrts}
\end{figure}

The measurements in the $t$-channel are performed using events with exactly one isolated lepton (electron or muon) and two or three jets. One of the jets is identified as $b$-jet. Additional cuts on kinematic observables are applied to further remove background. Results are available from ATLAS and CMS for center-of-mass energies $\sqrt{s}$ of both 7 TeV and 8 TeV. Figure~\ref{fig:singlet-sqrts} gives an overview of the results.

\begin{figure}[tb]
\centerline{
\epsfxsize=2.5in\epsfysize=2.4in\epsfbox{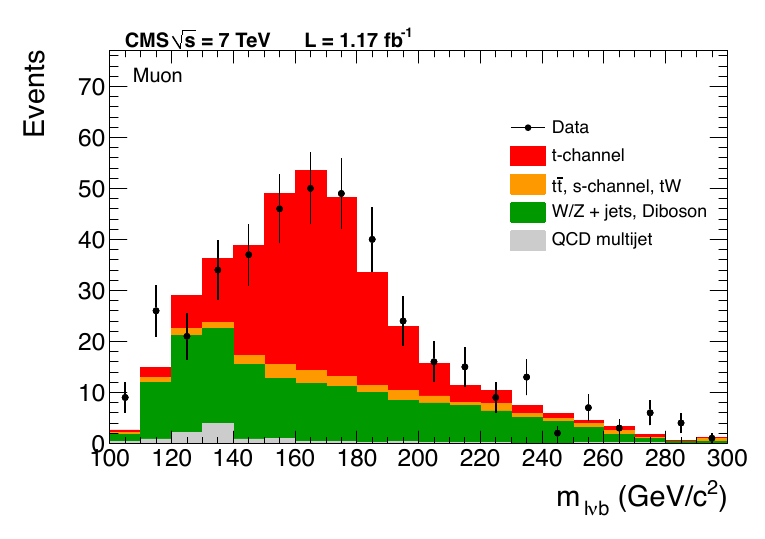} 
\epsfxsize=2.6in\epsfysize=2.3in\epsfbox{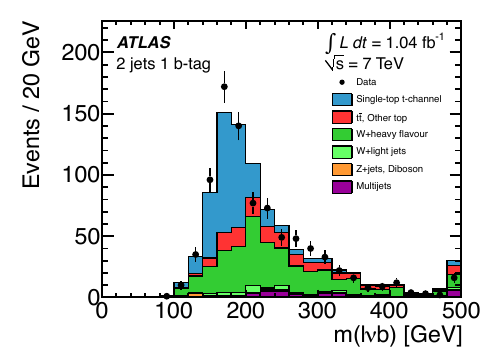} 
}
\caption[*]{Single top quark production in the $t$-channel at 7 TeV at the LHC. The distributions of invariant mass of the $b$-jet and the isolated lepton are displayed for CMS (left)~\cite{cms-top-11-021} and ATLAS (right)~\cite{atlas-t-channel}.}
\label{fig:lhc-singlet}
\end{figure}

In the CMS analysis of the dataset with $\sqrt{s}=7$ TeV, in the electron channel, the missing transverse energy is required to be larger than 35 GeV. For the muon channel, the transverse mass of the W-boson is required to be larger than 40 GeV. The cross section measurement is obtained from the combination of three different analyses, of which two make use of multi-variate techniques. In the third analysis, the signal is determined by a template fit to the rapidity distribution $\eta_{j'}$ of the (untagged) recoil jet using the event category with one lepton and two jets of which one is $b$-tagged. Other event categories are used to control the backgrounds. Figure~\ref{fig:lhc-singlet}(left) shows the distribution of the invariant mass of the lepton and the $b$-tagged jet for events with $|\eta_{j'}|>2.8$. A clear signal of top quark events is seen. The results from the three analyses are combined to yield a measured cross section of $\sigma_t=67.2\pm6.1$ pb~\cite{cms-top-11-021}. Dominant uncertainties come from statistical limitations, the modeling of signal and backgrounds, as well as the $b$-tagging. 

In the ATLAS analysis, events with one lepton and two or three jets are selected, if the missing transverse energy is larger than 25 GeV. The $t$-channel cross section is measured by fitting the distribution of a multivariate discriminant, constructed with a neural network, yielding a result $\sigma_t=83\pm4(stat.)^{+20}_{-19}(syst.)$ pb~\cite{atlas-t-channel}. Dominant uncertainties arise from model uncertainties, such as the ISR/FSR scale, as well as the $b$-jet identification efficiency. The result is cross-checked by an independent analysis using a cut-based selection. In Figure~\ref{fig:lhc-singlet}(right) the distribution of the invariant mass of the lepton and the $b$-tagged jet is shown for two-jet events with $|\eta_{j'}|>2.0$. Here, the scalar sum of lepton, jets and missing transverse energy, $H_T$, is required to be larger than 210 GeV.
ATLAS also measures the ratio $R_t$ between the cross sections for top and anti-top quark production in the $t$-channel. The result is $R_t=1.81\pm 0.10(stat.)^{+0.21}_{-0.20}(syst.)$~\cite{atlas-t-channelr}. The ratio is sensitive to the ratio of up-quark and down-quark parton distribution functions of the proton. In the measurement of the ratio, the uncertainties common to both channels cancel.

Both ATLAS and CMS already presented $t$-channel cross section measurements using the data recorded in 2012 at a centre-of-mass energy of 8 TeV\cite{ATLAS-CONF-2012-132,cms-top-12-011}. In these analyses, the selection criteria are somewhat tightened with respect to the 7 TeV analyses described above, in order to cope with more severe backgrounds due to the increase of both pile-up events and center-of-mass energy. The results of the measurements are included in Figure~\ref{fig:singlet-sqrts}.

\begin{figure}[htb]
\centerline{
\epsfxsize=2.5in\epsfysize=2.4in\epsfbox{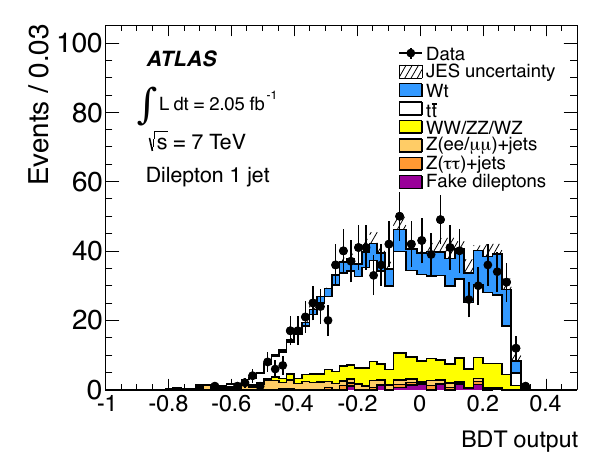} ~
\epsfxsize=2.5in\epsfysize=2.3in\epsfbox{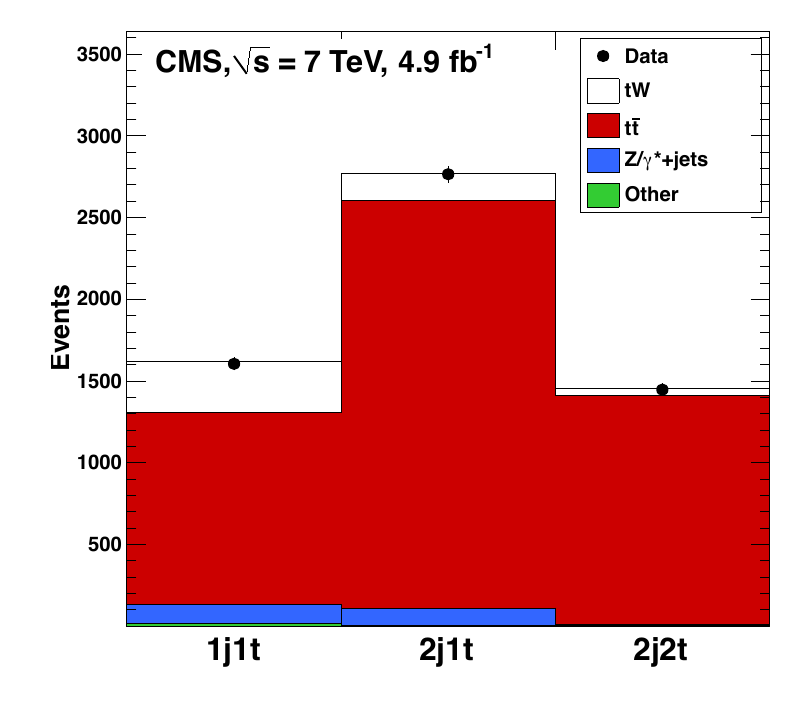} 
}
\caption[*]{Single top quark production in the $tW$-channel at 7 TeV at the LHC. The output distribution from the boosted-decision tree of the ATLAS analysis (left)~\cite{atlas-tW-channel} and the distribution of jet and lepton multiplicities in the CMS analysis (right)~\cite{cms-top-11-022}.}
\label{fig:lhc-singletW}
\end{figure}

The single top cross section in the $tW$-channel, inaccessibly small at the Tevatron, is sizable at the LHC. Theoretical calculations predict the cross section to be $\sigma_{tW}=15.7^{+1.3}_{-1.4}$pb~\cite{kidonakistw}. 
Single top events in the $tW$ channel form an important background to Higgs searches in the decay to two $W$ bosons. 
The ATLAS and CMS experiments have performed first measurements of the single top cross section and observe signals at significances of 3.3 and 4.0 standard deviations, respectively~\cite{atlas-tW-channel,cms-top-11-022}. 
Events with two leptons (electron or muon) and at least one jet are selected. In addition, the missing transverse energy in the event is required to be larger than 50 (30) GeV in ATLAS (CMS). Events in the same-flavour channels (with two electrons or two muons) are rejected if the invariant mass is between 81 GeV and 101 GeV, thereby removing backgrounds from Z+jet events. In the CMS analysis, the jet is required to be $b$-tagged. 
The dominant background originates from top quark pair production where both $W$ bosons from the top quarks decay leptonically.
Both ATLAS and CMS extract the cross section using a template fit to a boosted decision tree (BDT) discriminant distribution. In Figure~\ref{fig:lhc-singletW}(left) the distribution of the BDT discriminant is shown for the event category with two leptons and one jet. The ATLAS measurement yields $\sigma_{tW}=16.8\pm 2.9 (stat)\pm4.9 (syst)$pb~\cite{atlas-tW-channel}. The cross section as measured by CMS is $\sigma_{tW}=16^{+5}_{-4}$pb~\cite{cms-top-11-022}, in good agreement with the Standard Model expectation.
In Figure~\ref{fig:lhc-singletW}(right) the distribution of jet and lepton multiplicities in a cut-based cross check analysis, as performed by CMS, is displayed. A clear signal is seen in the two lepton and one jet category.
 
No evidence has been reported so far for single top production in the $s$-channel. The Standard Model cross section prediction is $\sigma_s=4.6 \pm 0.3$ pb~\cite{kidonakiss}. The event signature in this channel consists of one lepton and jets, very similar to that for the top quark pair, QCD and W+jet events.
ATLAS reports an analysis in which an upper limit on the $s$-channel production cross section of $\sigma_s<26.5(20.5)$ pb observed (expected) is set~\cite{atlas-s-channel}.

\section{Conclusions}
Top quark production is a field of rapid progress. The measurements provide important information about the production process as described in QCD, as well as sensitivity to possible new physics. 
At the Tevatron, precise final results are becoming available. At the LHC, large statistics and center-of-mass energy give access to a new realm of top quark precision physics.
Measurements of inclusive cross sections have been performed for both top quark pair and single top production processes, exploiting all accessible decay channels. For top quark pair production, precise differential cross section results as well as studies of jet multiplicity distributions are available. In the sector of single top production, evidence for the associate production of top quarks with a $W$ boson is reported for the first time.
All measurements are in good agrement with Standard Model expectations.

\section*{Acknowledgements} 
I would like to thank the organizers of the PIC2012 conference for the realization of a very 
useful and enjoyable conference.

\end{document}